\documentclass[sigplan]{acmart}

\usepackage{hyperref}
\newcommand{\myhy}[2]{\hyperref[#1]{\color{green}\setulcolor{red}\ul{#2}}}
\hypersetup{%
	colorlinks=true,% hyperlinks will be coloured
	linkcolor=green,% hyperlink text will be green
	linkbordercolor=red,% hyperlink border will be red
}

\usepackage[utf8]{inputenc}
\usepackage{times}
\usepackage{color}
\usepackage[noend]{algorithmic}
\usepackage{algorithm}
\usepackage{graphicx}
\usepackage{subfig}
\usepackage{pgf}
\usepackage{tikz}
\usetikzlibrary{arrows,automata,shapes,shadows,calc}
\usetikzlibrary{positioning}
\usepackage[]{todonotes}
\usepackage{varwidth}
\usepackage{xspace}
\usepackage{tabularx}
\usepackage{colortbl}
\usepackage{nicefrac}
\usepackage{stackengine}
\usepackage{setspace}

\usepackage{float}

\usepackage{listings}
\usepackage{lstautogobble}
\usepackage{textcomp}

\usepackage{graphicx}
\graphicspath{{./images/}}

\usepackage{placeins}

\usepackage{array}

\newcommand*{\fullref}[1]{\hyperref[{#1}]{\ref*{#1} \nameref*{#1}}}

\lstdefinelanguage[ECMAScript2015]{JavaScript}[]{JavaScript}{
	morekeywords=[1]{await, async, case, catch, class, const, default, do,
		enum, export, extends, finally, from, implements, import, instanceof,
		let, static, super, switch, throw, try},
	morestring=[b]` % Interpolation strings.
}
\lstdefinelanguage{JavaScript}{
	morekeywords=[1]{break, continue, delete, else, for, function, if, in,
		new, return, this, typeof, var, void, while, with},
	morekeywords=[2]{false, null, true, boolean, number, undefined,
		Array, Boolean, Date, Math, Number, String, Object},
	morekeywords=[3]{eval, parseInt, parseFloat, escape, unescape},
	sensitive,
	morecomment=[s]{/*}{*/},
	morecomment=[l]//,
	morecomment=[s]{/**}{*/}, % JavaDoc style comments
	morestring=[b]',
	morestring=[b]"
}[keywords, comments, strings]

\lstalias[]{ES6}[ECMAScript2015]{JavaScript}

\definecolor{mediumgray}{rgb}{0.3, 0.4, 0.4}
\definecolor{mediumblue}{rgb}{0.0, 0.0, 0.8}
\definecolor{forestgreen}{rgb}{0.13, 0.55, 0.13}
\definecolor{darkviolet}{rgb}{0.58, 0.0, 0.83}
\definecolor{royalblue}{rgb}{0.25, 0.41, 0.88}
\definecolor{crimson}{rgb}{0.86, 0.8, 0.24}

\lstdefinestyle{JSES6Base}{
	backgroundcolor=\color{white},
	basicstyle=\ttfamily,
	breakatwhitespace=false,
	breaklines=true`',
	columns=fullflexible,
	commentstyle=\color{mediumgray}\upshape,
	emph={},
	emphstyle=\color{crimson},
	extendedchars=true,  % requires inputenc
	fontadjust=true,
	frame=single,
	identifierstyle=\color{black},
	keepspaces=true,
	keywordstyle=\color{mediumblue},
	keywordstyle={[2]\color{darkviolet}},
	keywordstyle={[3]\color{royalblue}},
	numbers=left,
	numbersep=5pt,
	numberstyle=\tiny\color{black},
	rulecolor=\color{black},
	showlines=true,
	showspaces=false,
	showstringspaces=false,
	showtabs=false,
	stringstyle=\color{forestgreen},
	tabsize=2,
	upquote=true  % requires textcomp
}

\lstdefinestyle{JavaScript}{
	language=JavaScript,
	style=JSES6Base
}
\lstdefinestyle{ES6}{
	language=ES6,
	style=JSES6Base
}
\lstdefinestyle{json}{
	backgroundcolor=\color{white},
	basicstyle=\ttfamily,
	breakatwhitespace=false,
	breaklines=true`',
	columns=fullflexible,
	comment=[l]{:},
	commentstyle=\color{mediumgray}\upshape,
	emph={},
	emphstyle=\color{crimson},
	extendedchars=true,  % requires inputenc
	fontadjust=true,
	frame=single,
	identifierstyle=\color{black},
	keepspaces=true,
	numbers=left,
	numbersep=5pt,
	numberstyle=\tiny\color{black},
	rulecolor=\color{black},
	showlines=true,
	showspaces=false,
	showstringspaces=false,
	showtabs=false,
	string=[s]{"}{"},
	stringstyle=\color{royalblue},
	tabsize=2,
	upquote=true  % requires textcomp
}

\definecolor{GrayCodeBlock}{RGB}{241,241,241}
\definecolor{BlackText}{RGB}{110,107,94}
\definecolor{RedTypename}{RGB}{182,86,17}
\definecolor{GreenString}{RGB}{96,172,57}
\definecolor{PurpleKeyword}{RGB}{184,84,212}
\definecolor{GrayComment}{RGB}{170,170,170}
\definecolor{GoldDocumentation}{RGB}{180,165,45}
\newcommand{\remove}[1]{}

\newcommand{\squishlist}{
 \begin{list}{-}
  { \setlength{\itemsep}{0pt}
     \setlength{\parsep}{1pt}
     \setlength{\topsep}{1pt}
     \setlength{\partopsep}{0pt}
     \setlength{\leftmargin}{0.9em}
     \setlength{\labelwidth}{1.5em}
     \setlength{\labelsep}{0.4em} } }
\newcommand{\squishend}{
  \end{list}  }

\setcopyright{none}
\renewcommand\footnotetextcopyrightpermission[1]{}
\title{Towards a Transpiler for C/C++ to Safer Rust}
\author{Dhiren Tripuramallu}
\affiliation{%
	\institution{IIT Bhubaneswar, India}
	\country{}
}
\email{dtm11@iitbbs.ac.in }
\author{Swapnil Singh}
\affiliation{%
	\institution{IIT Bhubaneswar, India}
	\country{}
}
\email{ss93@iitbbs.ac.in}
\author{Shrirang Deshmukh}
\affiliation{%
	\institution{IIT Bhubaneswar, India}
	\country{}
}
\email{dsp13@iitbbs.ac.in }
\author{Srinivas Pinisetty}
\affiliation{%
	\institution{IIT Bhubaneswar, India}
	\country{}
}
\email{spinisetty@iitbbs.ac.in}
\author{Shinde Arjun Shivaji }
\affiliation{%
	\institution{Samsung R\&D Inst., Bangalore, India}
		\country{}
}
\email{shinde.arjun@samsung.com }
\author{Raja Balusamy}
\affiliation{%
	\institution{Samsung R\&D Inst., Bangalore, India}
	\country{}
}
\email{raja.balu@samsung.com }
\author{Ajaganna Bandeppa}
\affiliation{%
	\institution{Samsung R\&D Inst., Bangalore, India }
	\country{ }
}
\email{ajju.b@samsung.com}

\acmConference[ ]{}{}{}
\renewcommand{\shortauthors}{D. Tripuramallu et al.}

\begin{document}
\begin{abstract}
\vspace{-0.5em}
\textit{Rust} is a multi-paradigm programming language developed by Mozilla that focuses on performance and safety. 
%Rust is known for its advanced safe concurrency capabilities.  
Rust code is arguably known best for its speed and memory safety, a property essential while developing embedded systems. 
Thus, it becomes one of the alternatives when developing operating systems for embedded devices. How to convert an existing C++ code base to Rust is also gaining greater attention.
\remove{
Rust's syntax is similar to C++ but offers faster speed and memory safety. Rust code is arguably known best for its speed and memory safety, a property essential while developing embedded systems. Thus, it becomes one of the alternatives when developing operating systems for embedded devices.}
\remove{
\textit{Samsung Electronics} uses \textit{Tizen} as the operating system on its devices, such as Smartwatches and Smart TVs. Tizen, an operating system developed in C++, is directly responsible for interacting with the hardware. }
%Given Rust's memory safety and efficiency, Samsung intends to migrate Tizen from C++ to Rust.

In this work, we focus on the process of transpiling C++ code to a Rust codebase in a robust and safe manner. The manual transpilation process is carried out to understand the different constructs of the Rust language and how they correspond to C++ constructs. Based on the learning from the manual transpilation, a transpilation table is created to aid in future transpilation efforts and to develop an automated transpiler. We also studied the existing automated transpilers and identified the problems and inefficiencies they involved. The results of the transpilation process were closely monitored and evaluated, showing improved memory safety without compromising performance and reliability of the resulting codebase. The study concludes with a comprehensive analysis of the findings, an evaluation of the implications for future research, and recommendations for the same in this area. 
%The study highlights potential benefits of using Rust for low-level systems programming and its applications in other domains.
\end{abstract}

\keywords{Rust, C/C++, source-to-source compiler}
%\vspace{-0.5em}

\settopmatter{printfolios=true}

\maketitle

%%%%%%%%%%%%%%%%%%%%%%%%%%%%%
\vspace{-1.5em}
\section{Introduction}
%\subsection{Rust Language}
%\noindent
% \textit{Rust} is a system programming language that gives you access and fine-grained control over system resources, memory, etc. Anything that we can build with low-level C++ can be built with Rust. 
\textit{Rust} \cite{RustRef1,rustB1} is a systems programming language that provides developers with low-level access and control over system resources such as memory, file systems, and networking. This makes it ideal for building software applications that require performance, efficiency, and reliability, such as databases, compilers, and other low-level systems \cite{RustRef1,10.1145/3158154}.

Rust's notable feature is its absence of a garbage collector, distinguishing it from most other high-level languages. Rust achieves memory safety through its sophisticated type system that tracks variable lifetimes at compile-time. This enables Rust to automatically insert optimized LLVM/assembly instructions for memory deallocation, resulting in improved performance and stability \cite{rustB1}. 
%
%
%One of the unique features of Rust is that it does not have a garbage collector, which is a common feature in other high-level programming languages. This means that Rust automatically manages memory allocation and deallocation to prevent memory leaks. In addition, it also provides complete control over memory allocation and deallocation whenever required, allowing developers to optimize memory usage. All this, in turn, provides a performance boost and improved stability compared to languages with a garbage collector \cite{rustB1}.
%\newline
In addition to its low-level control and performance, the automatic memory allocation and deallocation also makes Rust a memory-safe language, which prevents common programming errors such as null pointer dereference and buffer overflows. 

\textbf{Why migrate to Rust?}: Given these features, it is easy to see why Rust is an attractive option for organizations that need to migrate their existing infrastructure to a new platform. By using Rust, organizations can take advantage of its low-level control, performance, and memory safety to build software that is faster, more efficient, and less prone to errors. 
%Whether you're building a new system from scratch or migrating an existing one, Rust is a powerful tool that can help you achieve your goals.

\remove{
	%\newline
	%\newline
	Rust was initially developed for the Mozilla Firefox browser to handle its memory leaks. However, its efficiency and advantages attracted many C++ developers who began to use Rust instead, commonly for game development.
	%\newline
	%\newline
}

Rust also has a strong focus on safety and security. The language enforces strict ownership and borrowing rules, making it difficult for developers to write code that can result in undefined behaviour, race conditions, and other security issues. This helps organizations reduce the risk of security vulnerabilities in their software and improve the overall reliability of their systems.
%\newline
%
Another important aspect of Rust is its highly optimized runtime system. The Rust compiler is designed to generate highly efficient machine code, making it a good choice for performance-critical systems. Rust also provides a number of performance-related features such as inline functions and zero-cost abstractions, which can help developers write high-performance code \cite{bugden2022rust} that is both readable and maintainable.
%\newline
%
%
%
\remove{In terms of community and support, Rust is a well-established and rapidly growing programming language. The language has a large and active community of developers who contribute to the language and its associated libraries, making it easier for organizations to find resources and support when they need it. Additionally, Rust has a rich ecosystem of libraries and tools that can help organizations quickly and easily build and deploy their systems.}
Given the range of benefits that Rust offers, it is an excellent choice that can help an organization achieve its goals and succeed in a rapidly changing technological landscape.
%\newline
\remove {
	In conclusion, Rust is a powerful, flexible, and secure systems programming language that provides organizations with a range of benefits for building high-performance, reliable, and safe software. Whether you're building a new system from scratch or migrating an existing one, Rust is an excellent choice that can help you achieve your goals and succeed in a rapidly changing technological landscape.
}

%\sr{Sec 1.1 may be shortened.. seems some points are repeated.. [UPDATED]} 
% \begin{enumerate}
	% \item Rust is syntactically similar to C++ but provides memory safety without garbage collection and increased speed.
	% \item It gives you low-level access to system resources
	% \item Rust specifically is designed to build software where performance and correctness are essential.
	% \end{enumerate}
\remove{
	\subsection{Migration of Samsung's Tizen operating system code base from C/C++ to Rust}
	Rust is specifically designed to build software where performance and correctness are essential. Properties that are essential while developing embedded systems. Thus making it one of the choices while developing embedded devices’ operating systems.
	\\
	\\
	\textit{Samsung Electronics} uses \textit{Tizen} as the operating system on its devices, such as Smartwatches and Smart TVs. Tizen is directly responsible for interacting with the hardware. The organization intends to migrate the existing code base of Tizen from C/C++ to Rust to take advantage of its increased speed and memory safety.
}
%\vspace{-0.5em}
%\subsection{Problem and Contributions}
%\noindent

\textbf{Problem and Contributions}:
Organizations such as \textit{Samsung Electronics} are interested in migrating their existing C++ codebases to Rust without starting from scratch. Therefore, a reliable auto transpiler for C++ to Rust is crucial. In our study, we focused on the \textit{Tizen OS}, which is utilized by Samsung Electronics in their smart devices. We extensively explored existing auto transpilers and evaluated their ability to generate safe Rust code, using both simple programs and a full-scale Tizen's gperf module. Our analysis included assessing the accuracy and reliability of the generated code, identifying strengths and weaknesses of each tool. Unfortunately, our findings revealed that the existing transpilers were inconsistent in producing reliable and safe Rust code.
%\sr{ADD a couple of lines of summary on this [UPDATED]!!}

\remove{
	The project involves collaboration between Samsung R\&D, Bangalore and aims to convert the Tizen platform modules from C++ to Rust in a safe and reliable manner. 
}
%Our study/analysis of the exting transpilers for C++ to Rust showed the need to re
Based on the study of the existing auto transpilers, there is a need to explore further on building a safe and reliable transpiler for C++ to Rust, that led to the following main contributions of this work.
%
%
%
%The ultimate goal of the project is to use an automated transpiler to transpile the codebase, making the process more efficient for future modules.
%\newline
%
%

In this work, the focus is on understanding the different constructs of Rust and how they correspond to the constructs of C++. A deep dive into the language and its features is conducted, including the correlation between these constructs and how they can be implemented in Rust. For example, we explored how Rust's concept of references and borrowing replaces C++'s raw pointers and null pointers.
%\newline
%

We then proceeded to manually transpile a module, the gperf module of Tizen platform (originally written in C++), as a proof of concept. This manual process helped us gain a deeper understanding of the language and its constructs. Based on the learning from this process, we have created a transpilation table. 
This transpilation table that resulted will guide future transpilation efforts and also form the basis for the creation of an automated transpiler. We then also compared and provided some insights on the performance of C++ vs. Rust on the gperf module.

\remove{
	The transpilation process and the resulting codebase will be closely monitored to ensure the desired results are achieved.}

%We would like to acknowledge the inputs from Mr. Raja Balusamy and Team Samsung R\&D, Bangalore. Their inputs were very helpful and constructive during the course of the project.

\vspace{-1em}
%%%%%%%%%%%%%%%%%%%%%%%%%%%%%%%%%%%%%%%
\section{Some key concepts of Rust}
\label{sec:concepts}
%%%%%%%%%%%%%%%%%%%%%%%%%%%%%%%%%%%%%%%
Rust is a programming language that is designed to prioritize memory safety. To achieve this, Rust provides several features that help prevent common programming errors that can lead to crashes and security vulnerabilities. Here are some of the fundamental but important to the domain memory safety features in Rust. Please refer to Rust's official website \cite{rustlang} for complete documentation on the features.
\begin{figure}[!]
	\begin{minipage}{0.5\textwidth}
		\begin{center}
			
		%	\scalebox{0.85}{
%		\centering{
			\begin{lstlisting}[style=ES6]
		fn main() {
			let v = vec![1, 2, 3];
			let v2 = v;
			println!("v[0] is: {}", v[0]);
			/* error:use of moved value:`v`
			println!("v[0] is:{}", v[0]);*/
		}
			\end{lstlisting}
			\vspace{-3em}
			\caption{Example of ownership in Rust}
	%		\vspace{-1em}
			\label{fig:ownership}
		\end{center}
			\end{minipage}
%		}
%	}
		\end{figure}
		%

%
%\vspace{-0.75em}
%\subsection{Ownership}
%%%%%%%%%%%%%%%%%%%%%%%%%%%%%%%%%%%%%%%%%
\begin{figure}[hbt!]
%	\scalebox{0.9}{
	\begin{lstlisting}[style=ES6]
		fn main () {
			fn take(v: Vec<i32>) {
				// Implementation 
			}
			let v = vec![1, 2, 3];
			take(v);
			println!("v[0] is: {}", v[0]);
			/* error: use of moved value: `v`
			println!("v[0] is: {}", v[0]);*/
		}
	\end{lstlisting}
	\vspace{-2.5em}
	\caption{Example of ownership using functions in Rust}
	\vspace{-1em}
	\label{fig:ownership2}
%}
\end{figure}
\vspace{-0.5em}
\paragraph{Ownership} Rust's ownership system ensures that there is always exactly one binding to any given resource. When we assign a value to another binding, we are actually transferring ownership of that value. This transfer is known as a move in Rust, and it helps to prevent common bugs such as use-after-free errors and data races.

In Figure \ref{fig:ownership}, a new vector \textbf{\textit{v}} is created and initialized with the values 1, 2, and 3. Then a new variable \textbf{\textit{v2}} is created and assigned \textbf{\textit{v}}. This means that ownership of the data in \textbf{\textit{v}} is transferred to \textbf{\textit{v2}} and \textbf{\textit{v}} can no longer be used. An attempt is made to access the first element of \textbf{\textit{v}} after it has been moved to \textbf{\textit{v2}}, but this results in a compile-time error because \textbf{\textit{v}} has been moved and is no longer available in the current scope.

The same error occurs in Figure \ref{fig:ownership2} when the ownership of \textbf{\textit{v}} is passed to function \textbf{\textit{take}}. The function takes ownership, and the original variable can no longer be used.

\remove{ 
	In Rust, every value has an owner and the scope of the value is tied to the scope of its owner. When a value goes out of scope, Rust automatically frees the memory it was using. This helps prevent memory leaks and makes it easier to reason about the lifetime of data.

\begin{figure}[hbt!]
	\begin{lstlisting}[style=ES6]
		fn main() {
			let x = 5; 	let y = x;
			println!("x = {}", x);  // Error: use of moved value 
		}
	\end{lstlisting}
	\vspace{-2.5em}
	\caption{Example of ownership and borrowing in Rust}
	\label{fig:ownership}
\end{figure}

In Figure \ref{fig:ownership}, the variable \textbf{\textit{x}} is created and assigned the value \textbf{\textit{5}}. Since \textbf{\textit{x}} is the owner of the value \textbf{\textit{5}}, it has the right to manage its memory. The variable \textbf{\textit{y}} is then created and assigned the value of \textbf{\textit{x}}. At this point, ownership of the value \textbf{\textit{5}} is transferred from \textbf{\textit{x}} to \textbf{\textit{y}}. When the print function referring to \textbf{\textit{x}} is called, the compiler detects that ownership of the value \textbf{\textit{5}} has been transferred to \textbf{\textit{y}} and that \textbf{\textit{x}} is no longer a valid variable. Therefore, the compiler reports an error, preventing us from using a moved value.}

%\sr{May refer to Figure ~\ref{fig:ownership} and add a couple of lines explaining the example.[UPDATED]}
%
\vspace{-0.5em}
%\subsection{Borrowing}
\paragraph{Borrowing}
It is allowed in Rust to borrow a value, which grants temporary access to the variable. Instead of taking variables as arguments in functions, references of the variables are taken as arguments. This way, ownership of the resource is borrowed rather than owned. There are 2 types of borrowing both important for the domain, namely Mutable Borrows, and Immutable Borrows.
\textbf{Rules of Borrowing}:
In Rust, borrowing rules state that a borrow must not outlast its owner's scope, and only one mutable reference or one or more references to a resource can exist at a time.  These rules prevent data races by ensuring that multiple pointers cannot access the same memory location at the same time, with at least one of them writing, without synchronization.

We discuss further about the concept of borrowing and the rules of borrowing in Rust via some examples in \href{https://github.com/CPPtoRust/CppToRustWork/blob/main/Paper_Appendix/Appendix.pdf}{\textit{\underline{Appendix B}}} available in the repository \cite{repo}.

\vspace{-0.75em}
\section{Related Work}
Rust, released in 2015, has consistently been named the "most loved programming language" for seven years in a row according to the Stack Overflow Developer Survey \cite{stackoverflow}. Its popularity has resulted in numerous research papers exploring its applications in software development and programming languages. The active Rust developer community also contributes to the abundance of online discussions, blog posts, and tutorials.

Studies have focused on migrating from languages like C/C++ to Rust due to its advantages in speed and safety, particularly for developing low-level systems. Open-source migration guides, such as "A Guide to Porting C/C++ to Rust," are available to assist developers in this process.

Martins \cite{martins2019benefits} investigated the benefits and drawbacks of using Rust in an existing codebase, specifically the EPICS framework. The author found that if EPICS had been written in Rust, approximately 41 bugs could have been easily identified and fixed. The study recommended selectively rewriting critical components in Rust, while maintaining a well-defined interface between the two languages. \remove{Rust was considered a suitable choice for new systems-level projects.}

Various automatic transpilation tools are available for converting C/C++ code to Rust, including C2Rust \cite{immunant}, CRust \cite{nishanthspshetty} and bindgen \cite{bindgen}. These tools offer different levels of functionality and limitations, which will be discussed in the subsequent sections of this paper.

Another study by Ling et al. \cite{ling2022rust} proposed a transpiler that transforms C code into safer Rust code by reducing the usage of the "unsafe" keyword. \remove{Their approach utilizes code structure pattern matching and transformation techniques to address "unsafe" usage in function qualifiers and unsafe blocks.}The authors' evaluation on open-source and commercial C projects demonstrated significantly higher ratios of safe code after the transformations.
Similarly, Lunnikivi et al. \cite{lunnikivi2020transpiling} discussed transpiling Python code to Rust to improve performance while maintaining readability. They proposed using Rust as an intermediate step for code optimization. The authors presented a transpilation process involving optional runtime types, the use of the pyrs \cite{konchunas} tool, manual refactoring, and validation testing. Their approach showed performance gains compared to accelerated Python implementations.

In the next section, we will explore the  effectiveness of the aforementioned transpilation tools, providing a comprehensive analysis of their features and limitations.

\vspace{-1em}

%%%%%%%%%%%%%%%%%%%%%%%%%%%%%%%%%%%%%%%
\section{C/C++ to Rust: On the existing tools}
%%%%%%%%%%%%%%%%%%%%%%%%%%%%%%%%%%%%%%%
In this section, we focus on evaluating the effectiveness of existing tools for transpiling C/C++ code to Rust. We begin by exploring the available tools and selecting a set of representative programs to test their transpilation capabilities. We then compare the resulting Rust code with the original C/C++ code to evaluate the quality of the transpilation. Through this evaluation, we aim to provide insights into the current state of C/C++ to Rust transpilation tools and their potential for adoption in real-world applications.

%\sr{We may add a few lines about the focus of this section. "We have initially explored the existing tools for transpilation of C/C++ to Rust. In this section, we report on the effectiveness of the existing tools by using a set of example C/C++ programs for the evaluation and comparison."[UPDATED]}

%\subsection{Examples considered and approach for analysis of the tools}
\vspace{-0.75em}
\subsection{Examples considered and approach}
To evaluate the performance of automated C/C++ to Rust transpilers, we selected various sets of C/C++ code that include common constructs in both languages. The goal was to analyze how well the transpiler can convert C/C++ code into memory-safe Rust code that performs the same task. The description of codes chosen for evaluation are given below and the exact codes are included in \href{https://github.com/CPPtoRust/CppToRustWork/tree/main/AutoTranspilation/Analysis_Examples}{\textit{\underline{Analysis\_Examples}}} available in the repository \cite{repo}.
\vspace{-0.5em}
\subsubsection{C Language}

\squishlist
	\item \textbf{Recursive Fibonacci Numbers}: This program (shown in \href{https://github.com/CPPtoRust/CppToRustWork/blob/main/AutoTranspilation/Analysis_Examples/C/Fibonacci.c}{\textit{\underline{Fibonacci.c}}} available in the repository \cite{repo}) was chosen to study how Rust code handles function calls, including self-calls and calls to other functions, when transpiled.
	
	\item \textbf{Linked List Implementation in C}: This program  
	(shown in \href{https://github.com/CPPtoRust/CppToRustWork/blob/main/AutoTranspilation/Analysis_Examples/C/LinkedList.c}{\textit{\underline{LinkedList.c}}} available in the repository \cite{repo}) utilizes structs and raw pointers, which have a distinct implementation in Rust. However, raw pointers are not encouraged in safe Rust code. Transpiling this program aids in comprehending how a transpiler converts structs and raw pointers.
	%\sr{We may add the program in an Appendix at the end of the paper.[UPDATED]}
\squishend
\vspace{-0.5em}
\subsubsection{C++ Language}
\squishlist
	
	\item \textbf{Catalan Numbers Problem}: This is a basic C++ code that uses simple constructs like loops (shown in \href{https://github.com/CPPtoRust/CppToRustWork/blob/main/AutoTranspilation/Analysis_Examples/C%2B%2B/CatalanNumbers.cpp}{\textit{\underline{CatalanNumbers.cpp}}} available in the repository \cite{repo}). The conversion of this program shows how these constructs are dealt with during the transpilation process.  %\sr{We may add the program in an Appendix at the end of the paper.[UPDATED]}
	\item \textbf{Basic Classes and Object Code}: This program (shown in \href{https://github.com/CPPtoRust/CppToRustWork/blob/main/AutoTranspilation/Analysis_Examples/C%2B%2B/BasicOOPs.cpp}{\textit{\underline{BasicOOPS.cpp}}} available in the repository \cite{repo}) is a basic implementation of Classes and Objects, i.e. OOP in C++. OOP is an essential part of Tizen development, and this program was selected to understand how classes and objects are converted during the transpilation process. 
	% \sr{We may add the program in an Appendix at the end of the paper.[UPDATED]}
\squishend
\vspace{-0.75em}
\subsection{Tools Explored}
This Subsection describes the automated transpilation tools we have identified and explored. 
Each tool has an overview along with its issues.
The full code transpilation results for each tool are available in \href{https://github.com/CPPtoRust/CppToRustWork/tree/main/AutoTranspilation/Analysis_Results}{\textit{\underline{Analysis\_Results}}} available in the repository \cite{repo}. %\sr{Is the appendix included?[UPDATED]}
\vspace{-0.75em}
\subsubsection{C2Rust} 

C2Rust \cite{immunant} is an open-source tool for converting C code to Rust code. It aims to automate the process of porting C code, reducing the time and effort required for manual translation while preserving the original behaviour and performance of the C code. However, there are some limitations to consider when using C2Rust.

%C2Rust is a powerful tool, but it has some limitations to consider:
C2Rust is a powerful tool, but has some limitations.
\squishlist
	
	\item \textbf{Complex C code}: C2Rust may struggle with complex C code that uses advanced language features, macros, or low-level system calls. The tool is imperfect and may generate incorrect or inefficient Rust code for such cases.
	
	\item \textbf{Dependencies}: C2Rust can only convert the C code itself, not any external dependencies or libraries. These dependencies must be ported manually or replaced with existing Rust libraries.
	
	\item \textbf{Performance}: The converted code may not perform as well as the original C code, especially if the C code is highly optimized. It is important to profile the converted code and optimize performance where necessary.
	
	\item \textbf{Manual intervention}: The conversion process with C2Rust is not perfect and may require manual intervention to correct errors or improve code quality. The converted code should be thoroughly tested and reviewed to ensure it behaves as expected.
	
	\item \textbf{Rust-specific concepts}: C2Rust may not always be able to translate complex C code into Rust code that follows best practices for the Rust programming language. The generated code may require manual modification to adhere to Rust conventions and idioms.
	
	\item \textbf{C99 support only}: C2Rust currently only supports C99 code, and there is no support for C++. This means that if the original code is written in a different dialect of C or in C++, it may not be able to be converted with C2Rust.
	
	\item \textbf{Unsafe code}: Since C2Rust primarily focuses on automating the conversion process, the generated Rust code may contain unsafe elements. This is because many of the unsafe constructs in Rust are used to match the behaviour of the C code. As a result, manual review and modification may be necessary to make the generated code safer and more idiomatic Rust.
	
\squishend

C2Rust is a tool designed to convert C code into Rust. Although it can be a valuable asset for porting C code, there are some important limitations to keep in mind. Firstly, C2Rust currently only supports C and does not have the capability to handle C++ code. Secondly, the converted code in Rust is unsafe, meaning that the user does not get the advantage of memory safety and increased speed by directly using the code. To use these benefits, the user needs to put in significant effort, roughly 90-95\%, into converting the output code to safe Rust. This requires a thorough understanding of Rust programming and a strong attention to detail to ensure that the code is safe and secure. Despite these limitations, C2Rust can still be a useful tool for those looking to port their C code to Rust, but it should be used with caution and with a deep understanding of the consequences of converting unsafe code.

\vspace{-0.75em}
\subsubsection{bindgen}
Rust bindgen \cite{bindgen} is a tool that automates the process of creating Foreign Function Interface (FFI) bindings between Rust and C/C++ libraries. It allows developers to access C libraries directly in their Rust code by creating bindings that match the functions and data structures of the C code.
\remove{
	
	For example, given the C header of \textbf{\textit{doggo.h}} in Figure \ref{fig:es6-route-api1}, bindgen produces Rust FFI code in Figure \ref{fig:es6-route-api2} allowing us to call into the doggo library's functions and use its types.
	
}
Thus, it is important to note that Rust bindgen is not a transpiler; it generates FFI bindings for C/C++ code. It creates bindings that enable Rust to interact with the C/C++ code but does not convert the code into Rust. Therefore, the safety of these bindings relies on the memory safety of the original C/C++ code, and Rust bindgen does not add any additional safety measures.

%%%%%%%%%%%%%%%%%%%%%%%%%%%%%%%%%%%%%%%%%%%%%%%%%%%%
\begin{figure}[t]
	\begin{lstlisting}[style=ES6]
		/** Crust doesn't resolve C/C++ dependencies or included header.
		* You may have to define your own module and implement those functionality in Rust
		* Or you can translate header file with Crust to produce Rust code. *
		* >>>>>>>> # include < bits / stdc ++ . h >
		**/
	\end{lstlisting}
	\vspace{-3em}
	\caption{Code excerpt from the transpilation}
	\label{fig:Crust1}
\end{figure}
%\vspace{-2em}
%

As a result, the transpilation of C/C++ code into Rust was not pursued in this context. The objective was to fully transpile the modules into Rust rather than simply relying on the C library. This approach ensures the codebase is fully converted and can take advantage of Rust's safety and performance benefits.

\begin{figure}[!]
	\begin{lstlisting}[style=ES6]
		// Importing Token Crate
		use crate::library::lexeme::token::Token;
	\end{lstlisting}
	\vspace{-3em}
	\caption{Crate import example}
	%	\vspace{-1em}
	\label{fig:Crust2}
\end{figure}

\vspace{-0.5em}
\subsubsection{CRust} 
CRust is an open-source project on GitHub created by a developer based in Bangalore \cite{nishanthspshetty}. It is designed as a typical compiler and involves lexical analysis and parsing of the input source code. Although CRust has been able to solve several issues faced by previous tools, it still has some limitations that need to be considered. The limitations of CRust include:

\squishlist
	\item \textbf{Header file conversion}: CRust is unable to convert included header files and resolve dependencies or header files in C/C++. As seen in Figure \ref{fig:Crust1} it provides message stating the same in the output file whenever it finds a header file in the source code. A workaround is to convert the header files separately and then include them using the Rust crate syntax as shown in Figure \ref{fig:Crust2}. 
	
	%\sr{Add a line refering to Figure \ref{fig:Crust1}[UPDATED]}

	\item \textbf{Unknown corresponding functions}: In cases where the transpiler does not know the corresponding functions in Rust, it directly copies the function from the C/C++ file. This will require manual correction or the use of a C library package. C Library functions \textit{printf} and \textit{getchar} in the source code of Figure \ref{fig:crust3} are copied as it is in Figure \ref{fig:crust4}, their Rust counterparts are not present in the transpiled code.
	
	%\sr{Refer to Figure \ref{fig:crust3} and \ref{fig:crust4} and briefly explain in a line.[UPDATED]}
	
	\begin{figure}[b]
		\begin{lstlisting}[style=ES6]
			int main(){
				int n = 9;
				printf("%d\n", fib(n));
				getchar();
				return 0; }
		\end{lstlisting}
	\vspace{-2.3em}
		\caption{Example C Code}
		\label{fig:crust3}
	\end{figure}
		\begin{figure}[b]
		\begin{lstlisting}[style=ES6]
			fn main() {
				let n: i32 = 9;
				printf("%d\n", fib(n));
				getchar();
				return 0; }
		\end{lstlisting}
		\vspace{-2.5em}
		\caption{Resultant Rust Code}
		\label{fig:crust4}
	\end{figure}
	
	\item \textbf{Class support}: CRust claims to support classes in C++, but its performance is not optimal. It may throw errors for index out of bounds, enter into an infinite loop during execution, or have other issues with its core logic for parsing class declarations.
	
	\item \textbf{Preprocessors}: There is code that supports preprocessors like HeaderDefine, HeaderInclude, HeaderIfDefineStart, and HeaderIfDefineEnd, but it is currently not in a working state. This can be extended by changing the code.

	\remove{
		\item \textbf{Pointer support}: Since Rust does not have memory pointers and relies on variable borrowing, CRust does not support programs with pointers in C/C++. Additionally, there is no support for functions that involve allocating memory, such as malloc and calloc.
	}
\squishend
\vspace{-0.5em}
\subsubsection{Verdict on tools}
The study explored the following tools for automatic transpilation of C/C++ code to Rust. 
\squishlist
	
	\item \textbf{C2Rust} - This tool supports only C, not C++, and the converted code is in unsafe Rust, which requires almost 90-95\% effort to be converted to safe Rust.
	
	\item \textbf{bindgen} - This tool creates a Foreign Function Interface to access the C/C++ code. It is not a transpiler and not suitable for the purpose of transpilation.
	
	\item \textbf{CRust} - This transpiler is suitable for basic C/C++ code but requires manual intervention for conversion to Rust. Class conversion is not supported, which limits its use case.
	
\squishend

We concluded that CRust seemed to be the most promising tool for converting small chunks of code and can be used for about 40\% of the transpilation work of basic code. However, if the issue of class transpilation can be fixed, it may provide a head-start in transpilation. Also, some tools (such as Corrode \cite{jameysharp}) could not be built and are no longer supported. Therefore, those tools were not evaluated.

\vspace{-0.5em}
\subsection{Transpilation of \textit{gperf} module using CRust}
\vspace{-0.5em}
%%%%%%%%%%%%%%%%%%%%%%%%%%%%%%%%%%%%%%%%%%%
To assess the feasibility of migrating Tizen modules to Rust, we attempted to transpile existing source files in the Tizen's \textit{gperf} module using the CRust tool. As previously stated, \textbf{CRust} is the only automatic transpilation tool that can be considered for converting C/C++ to safe Rust code. We compared the outcome of the automatic transpilation with that of the manual transpilation. The comparison was based on the percentage of lines of C/C++ code that were converted to equivalent Rust code. The goal was to determine the accuracy of the automatic transpilation performed by CRust and to evaluate how much manual intervention would be required.
%%%%%%%%%%%%%%%%%%%%%%%%%%%%%%%%%%%%%%%%
%\vspace{-0.75em}
%\subsubsection{Conversion Results}

%\textbf{Conversion Results:} Table~\ref{table:1} provides a summary of the conversion percentages achieved for various gperf files when transpiled from C/C++ to Rust using the CRust tool. The table includes information on the file name, conversion percentage, and remarks on the transpilation process. The full transpilation of the files is included in \href{https://github.com/CPPtoRust/CppToRustWork/tree/main/AutoTranspilation/CRust_Transpiled_gperf}{\textit{\underline{transpiled\_gperf}}} available in the repository \cite{repo}.

\textbf{Conversion Results:} 
The  \href{https://github.com/CPPtoRust/CppToRustWork/blob/main/Paper_CRust_gperf_transpilation_table/CRust_gperf_transpilation_table.pdf}{\textit{\underline{Conversion results table}}} available in our repository \cite{repo} presents the conversion results of each file in the gperf module.
%The table includes information on the file name, conversion percentage, and remarks on the transpilation process. 
The full transpilation of the files is included in \href{https://github.com/CPPtoRust/CppToRustWork/tree/main/AutoTranspilation/CRust_Transpiled_gperf}{\textit{\underline{transpiled\_gperf}}} available in the repository \cite{repo}.

%Appendix~\ref{sec:appendix:summaryTools} presents the conversion results of each file in the gperf module in table~\ref{table:1}, comparing the automatic transpilation using CRust with the manual transpilation. 
%\newline
%The results in general show that, the automatic transpilation required major manual intervention to complete the conversion process.

%\newline

The table shows that while some files could not be transpiled, others had conversion percentages ranging from 5\% to 45\%, with logic fragments of the code being successfully converted. However, some data types were not added to the variables, and some functions were ignored or copied as they are. Thus, it suggests that CRust can partially convert C/C++ code to Rust, but additional effort will be required to complete the conversion process.
%%%%%%%%%%%%%%%%%%%%%%%%%%%%%%%%%%%
\vspace{-1em}
\subsection{Summary- existing transpilation tools}
The CRust transpiler's performance fell short of expectations. It struggled with larger code sections,  failed to convert header files with class definitions into appropriate Rust equivalents. Furthermore, complications arose with .icc files, necessitating their conversion to .cpp format for analysis. The transpiler encountered difficulties in handling certain crucial source code files, leading to infinite loops. While it managed to convert smaller code snippets resembling C/C++ syntax, such as array access and arithmetic operations, some code was either disregarded or mistaken for comments due to issues with the comment delimiter. On a positive note, variable data types were accurately assigned in Rust. Overall, the CRust transpiler's outcome was disappointing, as much of the code was merely copied with transpiler comments signaling the need for further exploration of corresponding Rust translations.

Upon assessing the current state of automated transpilers, it becomes apparent that they cannot satisfactorily convert Tizen modules from C/C++ to Rust. The Tizen codebase heavily relies on Object-Oriented Programming (OOP) principles and extensively employs memory pointers and inline functions. The primary obstacle lies in the disparities between memory and class handling in C/C++ and Rust. Additionally, differences in syntax compound the challenges encountered during the transpilation process. These combined factors render existing transpilers ill-suited for successful conversion of Tizen modules.
\vspace{-0.5em}
%%%%%%%%%%%%%%%%%%%%%%%%%%%%%%%%%%%%%%%%%%%%%%%%%%%%%
\section{Mapping of C/C++ to Rust}
%%%%%%%%%%%%%%%%%%%%%%%%%%%%%%%%%%%%%%%%%%%%%%%%%%%%%
%\red{This section summarizes our study on the mapping of various constructs of C/C++ to Rust. }
We began by taking the gperf module of Tizen platform into consideration for manually transpiling the C++ files present inside it into the Rust codebase. The following are some of the fundamental, key and significant constructs of C++ with their equivalent implementations in Rust. And it's very important to note that the ideal transpilation would essentially be from any form (safe/unsafe) of C++ code to safe Rust. This includes memory-safe and unsafe C++ code. 
%\sr{Let us discuss..how we can better organize the content of this section !!}
%%%%%%%%%%%%%%%%%%%%%%%%%%%%%%%%%%%%%%%%
\vspace{-0.5em}
\subsection{Basic Constructs}
\vspace{-0.5em}
Let us see how some basic constructs in C++ such as global variables, and ternary operator are implemented in Rust.
\vspace{-0.5em}
\subsubsection{Global Variables}
The Rust language offers two ways, using const and static, and keywords.
\begin{figure}[hbt!]
	\begin{lstlisting}[style=ES6]
		const const_global: f32 = 2.4;
		static static_global: i32 = 10;
		static mut mut_global: i32 = 5;
		fn main() {
			unsafe{
				mut_global = mut_global + 1;
			}
		}
	\end{lstlisting}
	\vspace{-4em}
	\caption{Global Variables in Rust}
	\vspace{-1em}
	\label{fig:global}
\end{figure}
%			pub fn main() {
	%//Implementation}
%
%%%%%%%%%%%%%%%%%%%%%%%%%%%%%%%%%%%%%%%%%%%%%	
\squishlist
	\item \textbf{const/static without \textit{mut} keyword}: To declare immutable global variables.
	\item \textbf{static with \textit{mut} keyword}: To declare mutable global variables. However, in order to access the \textit{static mut} variable, we need to wrap it with the unsafe code. 
\squishend

	\remove{Note that a \textit{const} variable is constant that lives for the entire lifetime of a program and has no fixed address in memory. This is because it is effectively inlined to each place that it is used. A \textit{static} variable also lives for the entire lifetime of a program but unlike \textit{const}, it is not inlined up which means it is at a fixed location in memory.} 
	
	To avoid the use of \textbf{unsafe\{\}} code, it is better to use types from \textit{std::sync} module such as \textit{RwLock} or \textit{Mutex} that provide thread-safe ways to mutate shared state without the need for the unsafe keyword.The \textit{let} keyword is not permitted to be used in the global scope.
	
		The code snippet in Figure \ref{fig:global} defines an immutable constant \textbf{\textit{const\_global}} of type \textbf{\textit{f32}} with an initial value of \textbf{\textit{2.4}}, an immutable static variable \textbf{\textit{static\_global}} of type \textbf{\textit{i32}} with an initial value of \textbf{\textit{10}} and a mutable static variable \textbf{\textit{mut\_global}} of type \textbf{\textit{i32}} with an initial value of \textbf{\textit{5}}. \textbf{\textit{mut\_global}} is accessed in an \textit{unsafe} block.

	\remove{\subsubsection{Size}
	\squishlist
		\item Of a Data Type
		
		Use std::mem::size\_of::$\langle$data\_type$\rangle$() to get the size of a data type in bytes, as shown in Figure \ref{fig:dtypesize}. 	The code snippet in Figure \ref{fig:dtypesize} uses Rust's \textbf{\textit{std::mem}} module to obtain the size in bytes of the \textbf{\textit{i32}} data type, which represents a signed integer with a width of 32 bits. The result of the function call to \textbf{\textit{size\_of}} denotes the number of bytes required to store a single value of type \textbf{\textit{i32}}.
		\begin{figure}[hbt!]
			\begin{lstlisting}[style=ES6]
		let size_of_datatype = std::mem::size_of::<i32>();\end{lstlisting}
			\vspace{-2.0em}
			\caption{Size of a Data Type in Rust}
			\vspace{-0.5em}
			\label{fig:dtypesize}
		\end{figure}
		
		\item Of a Variable
		
		Use std::mem::size\_of\_val(\&variable) to get the size of a variable in bytes. The code snippet in Figure \ref{fig:varsize} creates an integer variable \textbf{\textit{x}} and initializes it to the value of \textbf{\textit{3}}. Then, the \textbf{\textit{std::mem::size\_of\_val}} function is used to determine the size in bytes of the variable \textbf{\textit{x}}.
		
		\begin{figure}[hbt!]
			\begin{lstlisting}[style=ES6]
		let x = 3;
		let size_of_variable = std::mem::size_of_val(&x);\end{lstlisting}
				\vspace{-2.0em}
			\caption{Size of a Variable in Rust}
				\vspace{-0.5em}
			\label{fig:varsize}
		\end{figure}
		\FloatBarrier
	\squishend}
		\begin{figure}[hbt!]
		\begin{lstlisting}[style=ES6]
			let a = if x > 5 {10} else {7};\end{lstlisting}
		\vspace{-2.0em}
		\caption{Ternary Operator Equivalent in Rust}
		\vspace{-1em}
		\label{fig:ternary}
	\end{figure}	
	\subsubsection{Ternary Operator}
	Unlike other languages, Rust does not have ternary operators. It uses the regular if-else construct. 	The code in Figure \ref{fig:ternary} is an example of a conditional expression in the Rust programming language. The variable \textbf{\textit{a}} is assigned a value based on the result of the condition \textbf{\textit{$x>5$}}. If \textbf{\textit{x}} is greater than \textbf{\textit{5}}, \textbf{\textit{a}} is assigned the value \textbf{\textit{10}}, otherwise it is assigned the value \textbf{\textit{7}}. This is a concise way of writing a basic if-else statement.
	\vspace{-0.5em}
%	\FloatBarrier
%	
	\subsubsection{Vector}
	A contiguous growable array type: Vec$\langle$T$\rangle$.
	%\squishlist
	\paragraph{Using Vec::new() method}
		The Rust code in Figure \ref{fig:vectornew} demonstrates two ways of creating a vector (a dynamic array) of signed 32-bit integers. The first method creates an empty vector using \textbf{\textit{Vec::new()}} and then inserts the integer \textbf{\textit{1}} into it using \textbf{\textit{vec.push(1)}}. The second  creates an empty vector with a pre-allocated \textbf{\textit{capacity of 5}} using \textbf{\textit{Vec::with\_capacity(5)}}. This can be useful if the program knows the expected size of the vector in advance to avoid costly reallocations as elements are added to the vector later on.
		\begin{figure}[hbt!]
				\vspace{-0.5em}
		\begin{lstlisting}[style=ES6]
		let mut vec:Vec<i32>=Vec::new();
		vec.push(1);
		//Construct empty vector with certain capacity
		let mut vec: Vec<i32> = Vec::with_capacity(5);
		\end{lstlisting}
		\vspace{-2em}
		\caption{Vector in Rust using new() method}
		\vspace{-1em}
		\label{fig:vectornew}
	\end{figure}

		%\item
							\begin{figure}[hbt!]
			\begin{lstlisting}[style=ES6]
				let v = vec![0, 2, 4, 6];
				println!("{}", v.len()); 
				//Prints the number of elements in the vector
				//Loop to iterate over the vector
				for i in v {
					//iterating through i on the vector
					print!("{} ",i);  }
			\end{lstlisting}
			\vspace{-2.5em}
			\caption{Vector in Rust using vec! macro}
			\vspace{-1em}
			\label{fig:vectorvec}
		\end{figure}
		\paragraph{Using vec! macro –}
		The Rust code in Figure~\ref{fig:vectorvec} defines a vector \textbf{\textit{v}} containing the values \textbf{\textit{[0, 2, 4, 6]}}. The first block of code prints the number of elements in the vector using the \textbf{\textit{len()}} method of the vector. The second block of code uses a for loop to iterate over the elements of the vector. The loop variable \textbf{\textit{i}} takes on the value of each element in the vector \textbf{\textit{v}} in turn, allowing the loop body to perform some operation on each element. In this case, the loop body simply prints each element separated by a space.

%	\FloatBarrier
%	\squishend
%%%%%%%%%%%%%%%%%%%%%%%%%%%%%%%%%%%%%%	
	\subsubsection{Do-While Loops}
	
	We can implement it using a regular loop block with a condition to break at the end. The Rust code in Figure \ref{fig:dowhile} demonstrates a loop construct, that is an infinite loop, that repeatedly calls a function \textbf{\textit{doStuff()}} as long as a condition, \textbf{\textit{c}} is met.  
	\begin{figure}[hbt!]
		\vspace{-0.5em}
		\begin{lstlisting}[style=ES6]
			loop {
				doStuff();
				if !c { break; }
			}\end{lstlisting}
		\vspace{-2.5em}
		\caption{Loop in Rust}
		\label{fig:dowhile}
	\end{figure}

%%%%%%%%%%%%%%%%%%%%%%%%%%%%%%%%%%%%%%%%%%%%%%%%%
		\vspace{-1em}
	\subsection{Non-Trivial Constructs}
%%%%%%%%%%%%%%%%%%%%%%%%%%%%%%%%%%%%%%%%%%%%%%%%%
	In this section, we show how some of the fundamental, yet non-trivial constructs, such as functions, classes and pointers in C++ are handled in Rust along with emphasizing on some key concepts in them. 
	\subsubsection{Functions}
	\squishlist
		\item Functions in C/C++ can be split into their declarations and their implementations as shown in Figure \ref{fig:functioncpp}. 
		\begin{figure}[htb!]
			\vspace{-0.25em}
			\begin{lstlisting}[style=ES6]
		// Declaration
		int foo(bool parameter1, const std::string &parameter2);
		// Implementation
		int foo(bool parameter1, const std::string &parameter2) {
			return 1; }
				\end{lstlisting}
			\vspace{-3em}
			\caption{Function Declaration and Definition in C++}
%			\vspace{-1em}
			\label{fig:functioncpp}
		\end{figure}
		%\FloatBarrier
		
		However, Rust does not allow such distinctions, so both of them has to be put in the same place as shown in Figure \ref{fig:functionrust}.  
		
		The Rust code in Figure \ref{fig:functionrust} defines a function foo that takes two parameters: a boolean value \textbf{\textit{parameter1}} and a reference to a string slice \textbf{\textit{parameter2}}, and returns an integer value of \textbf{\textit{1}}.
		
		\item	Rust also does not support function overloading. So to implement such traits in Rust, we can declare functions with different names and their respective parameter configurations as shown in Figure \ref{fig:functionoverloading}.
				\begin{figure}[hbt!]
					\vspace{-0.5em}
			\begin{lstlisting}[style=ES6]
	fn foo(parameter1: bool, parameter2: &str)->i32 {
		// implementation
		1 }
	\end{lstlisting}
\vspace{-3em}
			\caption{Function in Rust}
		%	\vspace{-1em}
			\label{fig:functionrust}
		\end{figure}
		\FloatBarrier
		\begin{figure}[hbt!]
			\vspace{-1em}
\begin{lstlisting}[style=ES6]
	fn functAdd3(a: i32, b: i32, c: i32)->i32{
		// implementation
		a+b+c }
	fn functAdd2(a: i32, b: i32) -> i32{
		//implementation
		a+b }
	\end{lstlisting}
	\vspace{-3em}
	\caption{Function Overloading Alternative in Rust}
%		\vspace{-1em}
	\label{fig:functionoverloading}
\end{figure}
The Rust code in Figure \ref{fig:functionoverloading} defines two functions \textbf{\textit{functAdd3}} and \textbf{\textit{functAdd2}} that perform addition of integer values. The function \textbf{\textit{functAdd3}} takes three parameters \textbf{\textit{a}}, \textbf{\textit{b}}, and \textbf{\textit{c}}, all of type \textbf{\textit{i32}} (signed 32-bit integer) and returns the sum of the three numbers. In case of function \textbf{\textit{functAdd2}}, the implementation is similar to \textbf{\textit{functAdd3}}, but takes only two parameters and returns the sum of \textbf{\textit{a}} and \textbf{\textit{b}}.

		\item Inline functions in C++ as shown in Figure \ref{fig:functioninlinecpp} 	can be implemented in Rust using the $\#[inline]$ keyword as shown in Figure \ref{fig:functioninlinerust}.
		
		\begin{figure}[hbt!]
			\vspace{-0.5em}
			\begin{lstlisting}[style=ES6]
		inline void sort_char_set (unsigned int *base, int len){
			//Implementation
		}\end{lstlisting}
			\vspace{-2.5em}
			\caption{Inline Function in C++}
				\vspace{-1em}
			\label{fig:functioninlinecpp}
		\end{figure}
		\FloatBarrier
		\begin{figure}[hbt!]
			\begin{lstlisting}[style=ES6]
		#[inline]
		fn sort_char_set(base: *mut u32, len: i32) {
			//Implementation
		}\end{lstlisting}
			\vspace{-2.5em}
			\caption{Inline Function in Rust}
			%	\vspace{-1em}
			\label{fig:functioninlinerust}
		\end{figure}
		%	\vspace{-1em}
		%\FloatBarrier
		%\vspace{-2em}
		The Rust code in Figure \ref{fig:functioninlinerust} defines a function \textbf{\textit{sort\_char\_set}} that sorts an array of unsigned 32-bit integers in place. The \textbf{\textit{\#[inline]}} attribute is a hint to the compiler to optimize the function by inlining it at the call site, which can improve performance by reducing the overhead of function calls.
	\squishend
	%%%%%%%%%%%%%%%%%%%%%%%%%%%%%%%%%%%%%%%
	\subsubsection{Classes}
	\squishlist
		\item Implementation of class in C++ is shown in Figure \ref{fig:classcpp}.
		\begin{figure}[hbt!]
			\vspace{-0.5em}
			\begin{lstlisting}[style=ES6]
			Class C{
				int a;	int b;	int c;
				public:
				C(int x, int y, int z)
				{ //constructor
					a = x;	b = y;	c = z;
				}
				void method1(int x, int y)
				{	a = x;	b = y; }
				int method2()
				{ a = 0;	return c; }
			}
			//creating an object
			C object(x,y,z);\end{lstlisting}
			\vspace{-2em}
			\caption{Class in C++}
			\label{fig:classcpp}
		\end{figure}
		Classes in Rust can be implemented using \textit{structs}. A separate \textit{impl} block is required in order to specify the struct methods. The above class can be implemented in Rust as shown in Figure \ref{fig:structrust}.

		The code in Figure \ref{fig:structrust} defines a struct \textbf{\textit{C}} with three fields \textbf{\textit{a}}, \textbf{\textit{b}}, and \textbf{\textit{c}}, which are of type \textbf{\textit{i32}}. The \textbf{\textit{pub}} keyword is used to make these fields public, which means they can be accessed from outside the struct. Two member functions \textbf{\textit{method1}} and \textbf{\textit{method2}} are defined for \textbf{\textit{C}}  inside the \textbf{\textit{impl}} block. \textbf{\textit{method1}} takes two integer parameters \textbf{\textit{x}} and \textbf{\textit{y}}, and updates the values of \textbf{\textit{a}} and \textbf{\textit{b}} using the self reference. \textbf{\textit{method2}} sets the value of \textbf{\textit{a}} to \textbf{\textit{0}} and returns the value of \textbf{\textit{c}}. The \textbf{\textit{\&mut self}} reference is used to indicate that \textbf{\textit{method1}} and \textbf{\textit{method2}} modify the fields of the struct, and they take ownership of the struct while they execute. A new object of type \textbf{\textit{C}} is created using the struct initialization syntax \textbf{\textit{C \{a:x, b:y, c:z\}}}, where \textbf{\textit{x}}, \textbf{\textit{y}}, and \textbf{\textit{z}} are the values of the constructor parameters.
		\begin{figure}[hbt!]
			\vspace{-0.5em}
			\begin{lstlisting}[style=ES6]
	pub struct C {
		a : i32,  b : i32,	c : i32,
   	}
	impl C {
		pub fn method1(&mut self, x:i32, y:i32)
		{
			self.a = x;	self.b = y; 
		}
		pub fn method2(&mut self) -> i32
		{
			self.a = 0;  self.c
		}
	}
	//constructor creating object
	let object = C {a:x, b:y, c:z};\end{lstlisting}
	\vspace{-2em}
			\caption{Struct in Rust}
			%	\vspace{-1em}
			\label{fig:structrust}
		\end{figure}
		\FloatBarrier
		
		\item Important thing to note here is that Rust does not support inheritance. Consider the following simple case of inheritance in C++ as shown in Figure \ref{fig:inheritancecpp}.
		
		\begin{figure}[hbt!]
			\vspace{-0.5em}
			\begin{lstlisting}[style=ES6]
			struct St1
			{
				int a;  int b;
			}
			struct St2 : St1
			{ 	int c;	}
			\end{lstlisting}
		\vspace{-2.5em}
			\caption{Inheritance in C++}
		%\vspace{-1em}
			\label{fig:inheritancecpp}
		\end{figure}
		%\FloatBarrier
		%
		Now to implement the same relation between two structs in Rust, we can either add a member of the parent struct, or we can simply add all the members of the parent struct and reinitialize them with the same values as of the parent.

		The code in Figure \ref{fig:inheritancerust} defines two structs, \textbf{\textit{St1}} and \textbf{\textit{St2}}, in Rust. \textbf{\textit{St1}} contains two fields of type \textbf{\textit{i32}}, named \textbf{\textit{a}} and \textbf{\textit{b}}. This struct has no parent or derived classes and it stands alone. \textbf{\textit{St2}} also contains two fields of type \textbf{\textit{i32}} named \textbf{\textit{a}} and \textbf{\textit{b}}, and an additional field \textbf{\textit{c}} of type \textbf{\textit{i32}}. The \textbf{\textit{c}} field is specific to \textbf{\textit{St2}} and has nothing to do with \textbf{\textit{St1}}.
		\\
		Alternatively, the second \textbf{\textit{St2}} struct definition shows how to define a struct with a parent using composition. It has a field named \textbf{\textit{parent}} of type \textbf{\textit{St1}}, which represents the parent struct. The \textbf{\textit{St2}} struct also has a unique field named \textbf{\textit{c}} of type \textbf{\textit{i32}}. By using composition, \textbf{\textit{St2}} inherits the fields of \textbf{\textit{St1}} through the \textbf{\textit{parent}} field. Also, this \textbf{\textit{parent}} field is not visible outside of the \textbf{\textit{St2}} struct.
		
		\begin{figure}[hbt!]
			\vspace{-0.5em}
			\begin{lstlisting}[style=ES6]
		struct St1
		{
			a : i32,   b : i32,
		}
		struct St2
		{
			a : i32,  b : i32,   c : i32,
		}
		//or
		struct St2
		{
			parent : St1,   c : i32,
		}\end{lstlisting}
			\vspace{-2.5em}
			\caption{Inheritance Alternatives in Rust}
				\vspace{-0.5em}
			\label{fig:inheritancerust}
		\end{figure}
	%	\FloatBarrier
	
		In Rust, inheritance is achieved through composition, where a struct can contain a field of another struct type, and the child struct can access the fields of the parent struct through the parent field.
		
	\squishend
	\subsubsection{Pointers and References}
	\squishlist
		\item As shown in Figure \ref{fig:pointercpp}, a pointer in most languages like C/C++ basically is the reference to the actual memory location of where the data is being stored. 	The same can be done in Rust as shown in Figure \ref{fig:pointerrust}.
		\begin{figure}[hbt!]
			\vspace{-0.5em}
			\begin{lstlisting}[style=ES6]
				int age = 27;
				int *age_ptr = &age;\end{lstlisting}
			\vspace{-2em}
			\caption{Pointer in C++}
			\vspace{-1em}
			\label{fig:pointercpp}
		\end{figure}
%		\FloatBarrier
		\begin{figure}[hbt!]
			\begin{lstlisting}[style=ES6]
	// This is a reference coerced to a const pointer
	let age: u16 = 27;
	let age_ptr: *const u16 = &age;
	// This is a mut reference coerced to a mutable pointer
	let mut total: u32 = 0;
	let total_ptr:*mut u32= &mut total;\end{lstlisting}
	\vspace{-2em}
			\caption{Raw Pointers in Rust}
		%	\vspace{-1em}
			\label{fig:pointerrust}
		\end{figure}
		\FloatBarrier
		
		The Rust code in Figure \ref{fig:pointerrust} demonstrates how to create pointers from references in Rust. The first example shows how to create a pointer from an immutable reference using the \textbf{\textit{*const}} keyword. The reference to \textbf{\textit{age}} is created using the \textbf{\textit{\&}} operator, and then it is coerced to a const pointer using the \textbf{\textit{*const}} keyword. The second example shows how to create a pointer from a mutable reference using the \textbf{\textit{*mut}} keyword. The mutable reference to \textbf{\textit{total}} is created using the \textbf{\textit{\&mut}} operator, and then it is coerced to a mutable pointer using the \textbf{\textit{*mut}} keyword.
		 		 
		The \textit{*const} keyword is used to specify that the pointer is immutable and cannot be used to mutate the data it points to. The \textit{*mut} keyword is used to specify that the pointer is mutable and can be used to mutate the data it points to.
				
		Note that most of the functions that we might want to use pointers in would be unsafe by definition. They must be inside an \textit{unsafe} block. Therefore, it is not recommended to use raw pointers in Rust. In Rust, a reference is also lifetime tracked by the compiler.
				
		\begin{figure}[hbt!]
			\vspace{-0.5em}
			\begin{lstlisting}[style=ES6]
			float *ptr = new float(10.25);\end{lstlisting}
			\vspace{-2em}
			\caption{Pointer in C++}
			\vspace{-0.5em}
			\label{fig:pointerrust2}
		\end{figure}
	%	\FloatBarrier
		
		%%%%%%%%%%%%
		\item In Rust, Box$\langle$T$\rangle$ is a smart pointer that can be used to allocate things on the heap similar to \textit{new} in C++. Basically, it is a type that provides ownership and lifetime management for heap-allocated values, and automatically deallocates the memory when it goes out of scope. The following pointer in C++ as shown in Figure \ref{fig:pointerrust2} 	can be implemented using Box pointers in Rust as shown in Figure \ref{fig:boxpointer}. 
		\begin{figure}[hbt!]
			\vspace{-0.5em}
			\begin{lstlisting}[style=ES6]
		let ptr:Box<f32> = Box::new(10.25);\end{lstlisting}
			\vspace{-3em}
			\caption{Box Pointer in Rust}
			\vspace{-0.5em}
			\label{fig:boxpointer}
		\end{figure}
		\FloatBarrier
		
		The Rust code in Figure \ref{fig:boxpointer} creates a \textbf{\textit{Box}} smart pointer that points to a heap-allocated \textbf{\textit{f32}} value initialized to \textbf{\textit{10.25}}. In this case, the Box pointer owns the \textbf{\textit{f32}} value and can be moved, but not copied, to other variables or functions.

		Raw pointers, i.e., $\ast const$ and $\ast mut$ can be obtained from Box pointer using Box::into\_raw() as shown in Figure \ref{fig:pointerconversion}.
		
		\begin{figure}[hbt!]
			\vspace{-0.5em}
			\begin{lstlisting}[style=ES6]
	let box_ptr:Box<int> = Box::new(5);	
	let raw_ptr:*mut i32 = Box::into_raw(box_ptr) as *mut i32;\end{lstlisting}
		\vspace{-2em}
			\caption{Box Pointer to Raw Pointer Conversion in Rust}
			\label{fig:pointerconversion}
		\end{figure}
		
		\item References with lifetime specifier; the main aim of lifetimes is to prevent dangling references, which cause a program to reference data other than the data it is intended to reference. Figure \ref{fig:references} shows various ways in which lifetime specifiers are specified to the references.
				
		\begin{figure}[hbt!]
			\vspace{-0.5em}
			\begin{lstlisting}[style=ES6]
			&i32        // a reference
			&'a i32     // a reference with an explicit lifetime
			&'a mut i32 // a mutable reference with an explicit lifetime\end{lstlisting}
			\vspace{-2em}
			\caption{References in Rust}
			\vspace{-0.5em}
			\label{fig:references}
		\end{figure}
		%\FloatBarrier
		
		Now, to use references of one struct as a member of another with lifetime specifiers, we can do as shown in Figure \ref{fig:structreference}.
		
		\begin{figure}[hbt!]
			\vspace{-0.5em}
			\begin{lstlisting}[style=ES6]
		struct PositionIterator<'a>
		{	_set :  &'a Positions, }
			\end{lstlisting}
		\vspace{-2.5em}
			\caption{Reference inside Struct in Rust}
			\vspace{-0.5em}
			\label{fig:structreference}
		\end{figure}
		\FloatBarrier
		
		The Rust code in Figure \ref{fig:structreference} defines a struct named \textbf{\textit{PositionIterator}} with a generic lifetime \textbf{\textit{'a}}. The \textbf{\textit{PositionIterator}} struct has one field named \textbf{\textit{\_set}} that is a reference to an instance of the \textbf{\textit{Positions}} struct. The lifetime specifier indicates that the \textbf{\textit{\_set}} reference is tied to the lifetime of the \textbf{\textit{PositionIterator}} struct. This code can be used to create an iterator over the positions in a \textbf{\textit{Positions}} object. The lifetime specifier is used to ensure that the iterator only has access to valid references to the \textbf{\textit{Positions}} object, and does not outlive it. By defining a lifetime on the \textbf{\textit{PositionIterator}} struct, Rust can check that the iterator's lifetime is valid and make sure that there are no dangling pointers or other memory errors. This helps ensure memory safety and avoid bugs in Rust code.
		\item NULL Pointers: there is no NULL in safe Rust unlike in C++. However, null pointers can be used in Rust along with raw pointers, which is unsafe to do so, using null()/null\_mut() functions in std::ptr.
		
		\begin{figure}[hbt!]
			\vspace{-0.5em}
			\begin{lstlisting}[style=ES6]
  use std::ptr;
  let p: *const i32 = ptr::null();
  let p_mut: *mut i32=ptr::null_mut(); 
 assert!(p.is_null()&&p_mut.is_null());\end{lstlisting}
	\vspace{-2.5em}
			\caption{Null Pointers (using raw pointers) in Rust}
			%	\vspace{-0.5em}
			\label{fig:nullpointerraw}
		\end{figure}
		\FloatBarrier
		
		In Figure \ref{fig:nullpointerraw}, a new immutable pointer \textbf{\textit{p}} and a new mutable pointer \textbf{\textit{p\_mut}} of types \textbf{\textit{*const i32}} and \textbf{\textit{*mut i32}} respectively are initialized it to null pointers using the \textbf{\textit{null()}} and \textbf{\textit{null\_mut()}} functions respectively from the \textbf{\textit{ptr}} module.
	
		In safe Rust, using enum \textit{Option} is the closest we get to using NULL. We can use the \textit{None} variant of it to represent no-value.
		
		\begin{figure}[hbt!]
			\begin{lstlisting}[style=ES6]
	let recipient: Option<&str> = None; 
	assert!(recipient.is_none());\end{lstlisting}
	\vspace{-2em}
			\caption{Option Enum in Rust}
			\label{fig:none}
		\end{figure}
		\FloatBarrier
	
	In Figure \ref{fig:none}, a variable \textbf{\textit{recipient}} is defined as an \textit{Option} of a string slice with an initial value of \textbf{\textit{None}}. The assert statement verifies that the \textbf{\textit{recipient}} variable is indeed \textbf{\textit{None}}.
	
	\squishend

%%%%%%%%%%%%%%%%%%%%%%%%%%%%%%%%%%%%%%%%%%%%%%%%%%%
\subsection{Summary and some observations: On understanding the mapping of C++ to Rust, and manual transpilation of the gperf module}
As can be seen, transpiling from C/C++ to Rust is not direct. Sometimes, it may require a totally different and indirect way of writing a piece of C/C++ code in Rust. Some features that are supported in one language may not be supported in the other, just like NULL pointer and inheritance concepts that are supported in C++ but not in Rust. 

Only a few topics are explained above. There are several other complex concepts such as traits, command line parsing, naming conventions, referencing, error handling, memory disposal, and visibility of fields and methods in structs that includes usage of \textit{pub} keyword and nested structs. 
Error handling using \textit{Result} enum is discussed in \href{https://github.com/CPPtoRust/CppToRustWork/blob/main/Paper_Appendix/Appendix.pdf}{\textit{\underline{Appendix A}}} available in the repository \cite{repo}.

%Error handling using \textit{Result} enum is discussed in Appendix A in the \textit{Appendix} PDF under \textit{manual\_transpilation} folder present in repository\cite{repo}.

Though there are C/C++ bindings that can be utilized in Rust, it is not advised to do so as they usually are compatible with unsafe Rust.

	The gperf module in C++ was structured as separate header (.h), inline definition (.icc), and implementation (.cc) files. However, in Rust, all code is grouped under a single implementation file (.rs), without any distinction between headers and implementations. Therefore, to translate a C++ module to Rust, we created a corresponding .rs file for each set of .h, .icc, and .cc files.
	
	\remove{Rust's module system is built around crates, which are equivalent to libraries or packages in other languages. A crate can contain multiple Rust modules, which can be imported and used using the `mod' and `use' keywords.}

	By organizing code into crates and modules, Rust provides a powerful and flexible way to structure code and manage dependencies. This allows developers to write maintainable and scalable code, while also ensuring that code can be reused across projects and shared with others.

\textbf{gperf} is a perfect hash function generator written in C++. It is used to generate the reserved keyword recognizer for lexical analyzers in several compilers and language processing tools \cite{gperf, schmidt_haible}. Overall, the gperf module is a complex system that requires careful management of memory and other system resources.
 \remove{\textbf{gperf} is a perfect hash function generator written in C++. It transforms an n element	user-specified keyword set W into a perfect hash function F. F uniquely maps keywords in W onto the range 0..k, where k >= n-1. If k = n-1 then F is a minimal perfect hash function. gperf generates a 0..k element static lookup table and a pair of C functions. These functions determine whether a given character string s occurs in W, using at most	one probe into the lookup table. It is used to generate the reserved keyword recognizer for lexical analyzers in several compilers and language processing tools \cite{gperf, schmidt_haible}.  
 
The gperf module consisted of components like the input, which reads in a set of keywords from an input file or from standard input, the output module direct the generated C code to a standard output. The option provides several options that permit users to trade-off execution time for storage space and vice versa. Users can expand the generated table size to produce a sparse search structure that generally yields faster searches, The search module performs hash table lookups using F. Overall, the gperf module is a complex system that requires careful management of memory and other system resources.}
	The gperf module implementation in C/C++ consists of almost 8500 lines of code \cite{gperf}. Using the understanding of the various programming constructs, we were able to manually convert the complete C/C++ code of the gperf module to Rust. The transpiled version of the codebase can be found in the refered repository \cite{repo}.  In the following section, we have briefly compared both the C/C++ and the Rust version of the gperf module implementations.   
	%%%%%%%%%%%%%%%%%%%%%%%%%%%%%%%%%%%%%%%%%%%%%%%%%%%

\vspace{-1em}
\section{Comparison of C/C++ to Rust (gperf)}
We performed a benchmark comparison between the runtime of C++ code and the corresponding Rust code of the files of the gperf module with some basic test cases on a machine with the following configuration:
\squishlist
\item Processor:	Intel(R) Core(TM) i7-7700HQ CPU @ 2.80GHz   2.80 GHz; Installed RAM:	16.0 GB (15.9 GB usable)
\item System type:	64-bit operating system, x64-based processor
\item Operating system: Windows 11 Home	
\squishend
\begin{table}[t]
	\scalebox{0.6}{
		\begin{tabular}{|c|c|c|c|c|}
			\hline
			\textbf{S.No} &
			\textbf{File Name} &
			\textbf{\begin{tabular}[c]{@{}c@{}}Rust(release) \\   (in milliseconds)\end{tabular}} &
			\textbf{\begin{tabular}[c]{@{}c@{}}C++ (no optimization) \\   (in milliseconds)\end{tabular}} &
			\textbf{\begin{tabular}[c]{@{}c@{}}C++ (-o1 optimization) \\   (in milliseconds)\end{tabular}} \\ \hline
			1 & bool-array    & 2  & 6.3  & 6.5    \\ \hline 
			2 & positions     & 2.5  & 8.5  & 1.6  \\ \hline
			3 & keyword       & 1.1  & 2.9  & 1.4  \\ \hline
			4 & keyword\_list & 0.5  & 1.1  & 1.5    \\ \hline
			5 & options       & 3.7  & 8.1  & 7.8    \\ \hline 
			6 & hash-table    & 2  & 8  & 2   \\ \hline
			
		\end{tabular}
	}
\vspace{1em}
	\caption{Benchmarking table: Comparison of the runtimes of various files in the gperf module in different modes. %\sr{the gperf module codebase [UPDATED]}
	\vspace{-3.5em}
	}
	\label{tab:my-table}
\end{table}
%%%%%%%%%%%%%%%%%%%%%%%%%%%%%%%%%%%%%%%%%%%
The Rust code was tested in debug mode and in release mode, with the release version being faster than the debug version. In contrast, the C++ code was tested with and without optimization flags (-o1 and -o2). The version without optimization flags was slower than Rust's release version. However, the C++ versions with optimization flags were almost similar to Rust's release version, with Rust still outperforming C++ by a few milliseconds in many cases. Table \ref{tab:my-table} shows the actual runtimes of each version where each runtime value was calculated as an average of runtimes obtained over 10 runs.

One of the drawbacks of using optimization flags in C++ is that it can sometimes lead to unexpected runtime errors and bugs. Additionally, using optimization flags can sometimes lead to longer compile times and larger executable files \cite{optimize_options}.

The results of this comparison suggest that Rust's performance is on par with C++. Here, the optimization flags in C++ can help achieve similar performance levels as Rust release compilation, albeit with slightly higher runtime and potential drawbacks. The research highlights the importance of selecting the right programming language for specific use cases, with Rust being a good choice for low-level systems programming, and C++ being suitable for other domains.

Overall, this benchmark comparison demonstrates the performance of Rust and emphasizes the need for continued exploration and development of Rust for low-level systems programming. 
In the process, it is also observed that there is a need to carefully consider the use of optimization flags in C++ to balance performance gains with potential drawbacks. 
\vspace{-1em}
\vspace{-1.2em}
\section{Discussion on Challenges Faced}
%\sr{we may move this content somewhere later..adding a discussion section etc..}
Some of the challenges that were faced in this  process were:
\vspace{-1em}
\squishlist
	\item Learning the language: Rust is difficult. It has a complex syntax and a steep learning curve. It is designed to uniquely solve some very challenging problems in programming.
	\item Understanding Rust constructs: Before we could begin the transpilation we had to have a thorough understanding of the different constructs of Rust and how they correspond to C++ constructs, which required significant time and effort.
	\item Manually transpiling modules: The process of manually transpiling was time-consuming and prone to errors, as it requires a deep understanding of both the source code and the target language.
	\item Creating a transpilation table: The process of creating a transpilation table based on the learning from manual transpilation was challenging, as it required us to identify and document key differences between C++ and Rust.
	\item Transpiling certain language-specific constructs:
		\squishlist
			\item \textbf{Header and source files:} In C++, header files contain the declarations of classes, functions, and variables that are defined in source files. However, Rust does not have a direct equivalent of header files. Our approach was to put the declaration and implementation of the class in one place, but this can lead to a large and unwieldy source file. Another approach is to use Rust modules to organize declarations and definitions, but this requires a significant reorganization of the codebase.
			\item \textbf{Friend class:} In C++, the friend class construct allows a class to grant access to its private members to another class. However, Rust does not have an equivalent construct. Our approach was to use the crate level visibility in Rust to allow a module to access the private members of another module, but this can lead to decreased encapsulation and increased coupling between modules.
			\item \textbf{Type inference:} C++ allows for implicit type conversions and coercion, which can result in unexpected behavior. Rust, on the other hand, is a strongly typed language that uses type inference to ensure type safety. This can make the transpilation process challenging, as the types used in the original C++ code may not be immediately obvious, and converting these types to Rust types can be error-prone.
			\item \textbf{Pointers and references:} C++ makes extensive use of pointers and references, which can lead to memory management issues such as dangling pointers and memory leaks. Rust, on the other hand, uses a borrow-checking system that ensures memory safety at compile-time. Converting pointers and references from C++ to Rust can be challenging, as the semantics of these constructs are different in the two languages.
			\item \textbf{Exceptions:} C++ has a built-in exception handling mechanism that allows for graceful error handling. Rust, on the other hand, uses a system of Result and Option types to handle errors. Converting exception handling code from C++ to Rust can be challenging, as the two systems have different semantics and error handling mechanisms.
		\squishend
%	\item Monitoring the results: The transpilation process and the resulting codebase was required to be closely monitored to ensure the desired performance improvements and increased reliability.
\squishend

\vspace{-1em}
\section{Conclusion}

In this work, we aimed to transpile C++ code to Rust in a robust and safe way. We began by exploring existing transpilation tools for C++ to Rust conversion. 
Our analysis of the current state of the existing tools for C++ to Rust showed that they do not provide adequate conversion.

Thus, we began to understand the mapping of various constructs of both the languages. For our study, we have considered a fragment of the Tizen OS (gperf module of the Tizen OS in C++) implemented in C++.  By manually transpiling various components of the considered C++ codebase to Rust, we have created a transpilation table/mapping based on the learning from manual transpilation of various C++ source files to Rust. Our learnings of the mappings of various constructs will help towards developing an auto transpiler from C++ to Rust.

In the process, the considered codebase is successfully completely manually transpiled in to Rust.  We then performed benchmark comparisons between the runtime of the C++ code (both debug and with optimization flag) and the corresponding Rust code (release mode). We observed that we achieved better memory safety based on Rust's type system without any compromise on the performance. 
%We also observed that the performance of the Rust version is on par with the C++ version.  

Note that manually transpiling C++ code to Rust is a time-consuming and laborious process. The conversion requires careful attention to detail, debugging, and testing, which can take significant time and resources. It is often more efficient to write the code from scratch or to use an automated transpiler. Our attempt of manually transpiling the considered codebase is only to carefully understand the mapping between various constructs of both the languages, to study the feasibility of developing an automated transpiler.
The next phase of our research will focus on developing an automated transpiler for C++ to Rust.

In conclusion, our research highlights the potential benefits of using Rust for system programming and demonstrates the feasibility of transpiling C++ code to Rust. We believe that with further development, Rust can become a valuable tool for creating high-performance and memory-safe software.

%%%%%%%%%%%%%%%%%%%%%%%%%%%%%
\vspace{-0.5em}
\bibliographystyle{ACM-Reference-Format}
\bibliography{biblio}
%%%%%%%%%%%%%%%%%%%%%%%%%%%%%%%%%%%%%%%%%%%
\appendix

\end{document}